\documentstyle[preprint,prl,aps,psfig]{revtex}
\tightenlines

\begin{document}
\draft 


\title{ First principles simulations of liquid Fe-S under Earth's core
conditions }

\author{ Dario Alf\`e and Michael J. Gillan }
 
\address{ Physics Department, Keele University, Keele, Staffordshire
ST5 5BG, U.K.}

\date{\today}
\maketitle

\begin{abstract} 
First principles electronic structure calculations, based upon density
functional theory within the generalized gradient approximation and
ultra-soft Vanderbilt pseudopotentials, have been used to simulate a
liquid alloy of iron and sulfur at Earth's core conditions.  We have
used a sulfur concentration of $\approx 12 \% $wt , in line with the maximum
recent estimates of the sulfur abundance in the Earth's outer
core. The analysis of the structural, dynamical and electronic
structure properties has been used to report on the effect of the
sulfur impurities on the behavior of the liquid. Although pure sulfur
is known to form chains in the liquid phase, we have not found any
tendency towards polymerization in our liquid simulation. Rather, a
net S-S repulsion is evident, and we propose an explanation for this
effect in terms of the electronic structure. The inspection of the
dynamical properties of the system suggests that the sulfur impurities
have a negligible effect on the viscosity of Earth's liquid core.

\end{abstract}

\pacs{PACS numbers: 
61.25.Mv,  
66.20.+d,  
66.30.Jt,  
71.15.Pd  
}

\narrowtext

\section{Introduction}

The Earth's liquid outer core consists mainly of molten iron, but its
density is about $10\%$ too low to be pure iron \cite{birch52}, so it
must contain also some light element. The nature of the light element
is still uncertain, and during the last forty-five years the main
proposed candidates have been carbon \cite{birch52,clark,urey},
silicon \cite{birch52,macdonald,ringwood59,ringwood61,ringwood66},
magnesium \cite{alder}, sulfur
\cite{clark,urey,birch64,mason,murthy,lewis}, oxygen
\cite{alder,birch64,ringwood77} or hydrogen
\cite{birch52,fukai,suzuki}. Due to motivations based on cosmic abundance,
models of Earth formation, and ability to dissolve into liquid iron
\cite{usselmanna,usselmannb,brett}, sulfur seems to be one of the most
likely light elements in the core. The properties of liquid iron and
iron alloys under very high pressures are of fundamental importance in
understanding the dynamics of the Earth's core, but they are difficult
to investigate because of the extreme conditions involved. A
particularly important property is the viscosity of the outer core,
since it determines the convective internal motions which are
responsible for the generation of the Earth's magnetic field.

First principles calculations have been shown to be very reliable for
the prediction of the structural and dynamical properties of a variety
of materials, including liquid metals \cite{stich,hafner1,hafner2}.
Since selenium and sulfur have very similar properties, it is relevant
to mention previous {\it ab initio} calculations of the structural,
dynamical, and electronic properties of liquid Ag-Se alloys
\cite{agse} and liquid Se \cite{se}, which have been shown to be in
very good agreement with experiments. Our own {\it ab initio}
calculations on pure liquid iron under Earth's core conditions have
demonstrated that the structure of the liquid is close packed, with a
coordination number $\ge 12$ and a diffusion coefficient of the same
order of magnitude as those of many liquid metals at ambient pressure
\cite{vocadlo,nature}.

We report here on a first principles investigation of the structural,
dynamical, and electronic structure properties of a liquid alloy of
iron and sulfur under Earth's core condition.  We have simulated a
liquid alloy with a $12 \%$ wt sulfur concentration, in line with the
maximum estimates for the sulfur abundance in the core
\cite{arhens}.

To our knowledge the only high pressure experimental work on a liquid
iron sulfur alloy is that of LeBlanc and Secco \cite{secco}. They have
studied a Fe$_{73}$S$_{27}$ (wt$\%$) (with the notation of the
original reference) liquid in a range of pressures between 2 and 5 GPa
and temperatures between 1100 and 1300$^{\rm o}$C, and found a value for the
viscosity about three order of magnitude higher than the ambient
pressure value. They have tentatively attributed this high viscosity
value to the formation of sulfur chains, or clusters. These aggregates
would impede the diffusion of the atoms in the liquid, resulting in an
enhancement of the viscosity.  Whether or not a similar sulfur effect
could be present also in the Earth's liquid core is a matter of
current dispute.  We remark that the temperatures studied by LeBlanc
and Secco are much lower and their pressures very much lower than
those in the Earth's core, so that it is not obvious that their
results have any relevance to the properties of the core.

The paper is organized as follows. In section \ref{method} we discuss
the theoretical framework, and in section \ref{solid} we present our
results for some solid Fe-S crystal structures, compared with 
other theoretical work. Then, in section \ref{liquid} we pass to the
discussion of the liquid, focusing our attention on the structural
(\ref{structure}), electronic (\ref{electronic}), and dynamical
properties (\ref{dynamics}). Finally we present our conclusions.

\section{Method}\label{method}

The first principles calculations presented here are based on density
functional theory within the generalized gradient approximation (GGA)
\cite{gga}. The electronic wave-functions are expanded in a plane-wave
basis set with a cut-off energy of 350 eV, and the electron-ion
interaction is described by means of ultrasoft Vanderbilt
pseudopotentials (PP)\cite{vanderbilt}, which allow one to use a much
lower number of plane-waves, comparing with a standard norm-conserving
PP, without affecting the accuracy of the calculations.  In the PP
approximation only the valence electrons are taken into account, while
the tightly bound core electrons are excluded from the
calculation. This approximation is usually perfectly justified, and
has been demonstrated to reproduce very well the all electron results
for transition metals. In particular, it has been accurately checked
for iron in our previous work \cite{vocadlo,nature}.  In spite of this
strong evidence, we considered it worthwhile to do some calculations
of the structural properties of solid FeS, and compare them with
all-electron full potential calculations of the same properties. The
results of these calculations are reported in the next section.

The iron PP is the same as that used in Refs. \cite{vocadlo,nature},
and has been constructed with a frozen [Ar] core and a $4s^13d^7$
reference valence configuration. The sulfur PP was constructed with
the [Ne] core and the $3s^23p^4$ reference configuration for the
valence states.  At the pressure conditions of the Earth's core the
distance among the atoms may become so small that the ionic cores 
overlap. This may result in a degradation of the PP approximation. The
iron PP has been constructed so as to minimize this problem, and its
quality has been checked elsewhere \cite{vocadlo,nature}.  The
reliability of the sulfur PP will be assessed in the next section.
Non-linear core corrections \cite{louie} are included throughout this
work.

The simulation of the liquid has been performed using {\it ab initio}
molecular dynamics (AIMD), with the forces calculated fully quantum
mechanically (within the GGA and the PP approximations), and the ions
moved according to the classical equation of motion.  We have used a
supercell approach with periodic boundary conditions. The first
pioneering work in AIMD was that of Car and Parrinello (CP)\cite{cp},
who proposed a unified scheme to calculate {\it ab initio} forces on
the ions and keep the electrons close to the Born-Oppenheimer surface
while the atoms move.  We have used here an alternative approach, in
which dynamics is performed by explicitly minimizing the electronic
free energy functional at each time step. This minimization is more
expensive then a single CP step, but the cost of the step is
compensated by the possibility of making longer time steps.  The
molecular dynamics simulations presented here have been performed
using VASP (Vienna ab initio simulation package). In VASP the
electronic ground state is calculated exactly (within a
self-consistent threshold) at each MD step, using an efficient
iterative matrix diagonalization scheme and a Pulay mixer
\cite{pulay}. Since we are interested in finite temperature
simulations, the electronic levels are occupied according to the Fermi
statistics corresponding to the temperature of the simulation. This
prescription also avoids problems with level crossing during the
self-consistent cycles. For more details of the VASP code see
Refs. \cite{kresse1,kresse2}.

Within this approach to AIMD it is important to provide a good
starting electronic charge density at each time step, so as to reduce the
number of iterations to achieve self-consistency. This is done usually
by a quadratic (or even multilinear) extrapolation of the charge. We
have used here a different scheme: at the beginning of each time step
the electronic charge density is extrapolated using the atomic charge
density and a quadratic extrapolation on the difference, i.e. the
charge is written as:
\begin{equation}
\rho(t) = \rho_{at}(t) + \delta \rho(t),
\end{equation}
where $\rho(t)$ is the self-consistent charge density at time $t$, and
$\rho_{at}(t)$ is the sum of the atomic charges. At time $t+dt$ the
charge is written as the sum of the atomic charges, which can be
calculated exactly and cheaply, and a quadratic extrapolation on
$\delta \rho$.  We have found that for liquid iron this scheme
provides a much better starting charge compared with a conventional
extrapolation of the whole charge, resulting in a reduction of CPU
time of almost a factor two.

\section{Solid FeS}\label{solid}

Solid FeS adopts a modified NiAs structure at zero pressure
\cite{wycoff}, and undergoes a first phase transition into a MnP
structure at 3.4 GPa \cite{prima} and a second transition into an
unknown structure at 6.7 GPa \cite{seconda1,seconda2,seconda3}. Mao
{\it et al. } have found FeS in an orthorhombic distorted B1 structure
at pressure above 11.5 GPa \cite{mao}.  Sherman \cite{sherman} has
done theoretical spin-unrestricted calculations on three possible
crystal structures. He has used the full potential linearized augmented
plane wave method (FLAPW) to study FeS in the NiAs(B8), CsCl(B2), and
NaCl(B1) crystal structures. He found that the CsCl structure is the
most stable at high pressure.

To confirm the accuracy of the pseudopotential approximation, we have
repeated the same calculations.  For each structure energy convergence
with respect to ${\bf k}$-points sampling has been checked.  We have
found total energies to be converged within 10 meV per atom by using
30, 20 and 10 ${\bf k}$-points in the irreducible Brillouin zone of
the FeS(B8), FeS(B2) and FeS(B1) structures respectively.  In the
FLAPW calculations the non-magnetic phase for the three sulfides was
found to be more stable then the magnetic one. We have found instead
that the magnetic phase is more stable for the B2 and the B8 compounds
(we have actually found that the B8 phase is anti-ferromagnetic),
while only the B1 phase shows no magnetic moment at all the volumes
investigated.  In Fig. \ref{fig:spin} we display the total energy as a
function of the volume for the B2 and the B8 structures compared with
the same calculations done in a spin-restricted scheme. The difference
in the total energy is clearly evident at low pressures. However, this
becomes negligibly small (even if never zero for the B2 structure) at
high pressure.  We want to point out that the disagreement regarding
the magnetism between our calculations and FLAPW calculations is
unlikely to be due to the PP approximation, as also previous
calculations on the structure of solid pure iron have shown to be in
very good agreement with all-electron calculations
\cite{vocadlo,nature}.  In Fig. \ref{fig:solid} we display
spin-unrestricted PP and FLAPW data, and in Tab. \ref{tab:solid} we
report the equilibrium density $\rho_0$, the bulk modulus $K$, and its
derivative with respect to pressure $K'$, as obtained from a fit of
the data to a Birch-Murnaghan equation of state, both for a
spin-restricted and a spin-unrestricted calculation. The FLAPW data
for the B8 and the B1 structures are not reported in
Ref. \cite{sherman} and have been deduced from a fit of the data to a
Birch-Murnaghan equation of state. Our calculated transition pressure
from the B8 to the B2 structure is 97 GPa, to be compared with 75 GPa
obtained in the FLAPW calculations.  Since the FLAPW results are
non-magnetic, the non-magnetic PP calculations have to be compared
with them.  Despite the slight difference in the equilibrium density,
the agreement between our non-magnetic calculations and FLAPW data is
good, and confirm the reliability of our PP calculations.

In order to check any possible effects at high pressure due to core
overlaps, we have repeated the calculations using a different sulfur
PP constructed with a shorter core radius (1.8 a.u. instead of 2.2
a.u.). We have not found any appreciable difference between the two,
and we have decided to use the PP having the large core radius for the
simulation of the liquid.

\section{The liquid}\label{liquid}

The possible amount of sulfur in the Earth's core is not certain, and
recent estimates provide a range from a few $\%$ wt to a maximum of
$\approx 10 \%$ wt \cite{arhens}. Since the effect of sulfur is likely
to be larger for larger concentrations, we decided to use the highest
possible amount of sulfur compatible with the current estimates. In
our simulations we have used 64 atoms in a cubic supercell. The
numbers of iron and sulfur atoms were 52 and 12 respectively,
resulting in $\approx 12 \%$ wt concentration and molar fraction of
$x_{Fe}=0.8125$ and $x_{S}=0.1875$.  In the liquid structure the
system is close packed. Since the majority of atoms are irons and
since in the hexagonal close packed structure solid iron is
non-magnetic at high pressure \cite{hcpiron}, we have used
spin-restricted calculations for all the liquid simulations. A
spin-unrestricted calculation on one configuration of the liquid has
confirmed that this is actually non-magnetic.

One of the possible effects of the impurities is the formation of
linear chains or small clusters. This fact could have important
effects on the transport properties of the whole liquid alloy, since
the impurity chains would impede the diffusion of the atoms, and
therefore would increase the viscosity.  In order to address this
possible effect, we decided to carry out two independent simulations,
starting with two very different atomic distribution
configurations. In the first case we have taken a previous pure liquid
iron simulation \cite{vocadlo} and substituted randomly iron with
sulfur; we will refer to this simulation as RS. In the second case we
have explicitly created a sulfur cluster, by transforming a chosen
iron atom together with 11 of its nearest neighbors into sulfur atoms;
we refer to this simulation as CS. The CS case has been performed to
give the sulfur atoms all the possible chances to stay together.  

Both the simulations have been done at a thermodynamic point
representative of the boundary between the Earth's solid inner core
and the liquid core. Here the temperature $T$ is uncertain; estimates
range from $4000$ to $8000$ K \cite{poirier}. The pressure is
accurately known, and it is 330 GPa \cite{poirier}.  In order to
compare the results of the present work with those obtained for pure
iron \cite{vocadlo,nature} we have used the same temperature $T=6000$
K. Since the sulfur has approximately the same size as iron at this
pressure, we argue that a small quantity of sulfur should not change
appreciably the pressure and therefore we have also used the same
volume per atom as in Refs. \cite{vocadlo,nature}.  This resulted in a
$\approx 8\% $ lower density, i.e. $\rho = 12.33$ g/cm$^3$ (for pure
iron it was $\rho = 13.30$ g/cm$^3$).  The Brillouin zone sampling has
been restricted to the $\Gamma$ point only, and the integration of the
classical equation of motion has been done using the Verlet algorithm
\cite{verlet}.  The temperature was controlled using a Nos\'e
thermostat \cite{nose,ditolla}. The quality of the simulation has been
checked by looking at the constant of motion; this usually shows a
drift which is due to a bad integration of the equation of motion
(time step too large) and/or a bad calculation of the forces. These
effects can both be easily controlled by acting on the time step and
on the self-consistency threshold on the electronic minimization,
which determines the accuracy with which the forces are
computed. However, a too short time step and/or a too small
self-consistency threshold may require a too expensive computational
effort, so one has to choose a judicious compromise.  We have used a
self-consistency threshold on the total energy of $1.5 \times 10^{-7}$
eV/atom and a time step of 1 fs; with these prescriptions the drift of
the constant of motion has been kept less then $\approx 7-8$ meV/atom
per ps.  We have simulated the RS system for 10 ps and the CS one for
3 ps. These simulations are continuations of a previous pure iron
simulation \cite{vocadlo,nature}, and then our starting configuration
would be an equilibrium configuration if we had only iron. Since we
suddenly transformed some iron atoms into sulfur atoms, this is not in
principle an equilibrium configuration for the new system. The time
needed to go from the equilibrium configuration of pure liquid iron to
that of the alloy is also an interesting quantity. For this reason no
equilibration time has been considered, and we report the whole
simulations starting from the very beginning.

\subsection{Structure}\label{structure}

The structural properties of the system have been inspected by looking
at the partial radial distribution functions (rdf), $g_{FeFe}(r)$,
$g_{FeS}(r)$, and $g_{SS}(r)$.  The partial rdf's are defined in such
a way that, sitting on one atom of the species $\alpha$, the
probability of finding one atom of the species $\beta$ in the
spherical shell $(r,r+dr)$ is $\rho_{\beta} 4\pi r^2 g_{\alpha
\beta}(r) dr$, where $\rho_{\beta} = x_\beta/V$ is the number density of the
species $\beta$ and $V$ is the volume per atom.

In Fig. \ref{fig:grRS} we display the rdf's calculated from 2.5 ps
time averages taken at four different starting times for the RS
simulation.  The four pictures provide a time analysis of the liquid
structure.  The two $g_{FeFe}(r)$ and $g_{FeS}(r)$ remain essentially
unchanged throughout the simulation.  Liquid sulfur is known to form
chains \cite{cotton}, the distance among the atoms for each pair being
$\approx 2$ \AA ~ at zero pressure.  If sulfur formed chains also in
the present case, this would result in a peak in the $g_{SS}(r)$ at
the position of the bond length. If sulfur just behaved as though it
was iron, then its partial rdf would be identical to the iron one. We
have not found either of the two behaviors. The form of the
$g_{SS}(r)$ clearly indicates that sulfur behavior is different from
iron, and at the same time it is also evident that sulfur atoms do not
form chains.  Rather, a S-S repulsion is suggested.  The same
indications come also from an inspection of the partial structure
factors (not reported). The analysis of the CS simulation is even more
interesting.  In Fig. \ref{fig:grCS} we display the rdf's for the
second simulation. In this case in the first panel of the figure we
display the rdf's averaged only over the first 0.5 ps of the
simulation. The reason for this short average time is that we have
found that the sulfur cluster dissociates quickly, and only in this
very short time can its existence be monitored. This is evident from
the presence of a peak in the $g_{SS}$ at $\approx 2$ \AA.  In the
second panel of the figure we display the partial rdf's averaged over
the last 2 ps of the simulation, i.e. starting the average 1 ps after
the beginning of the simulation. It is evident that the cluster has
completely dissociated and the rdf's have become essentially identical
to those of the RS simulation.  It is also interesting to compare the
structural properties of the alloy with those of the pure liquid
iron. In Fig. \ref{comparison} we display the iron rdf as calculated
in Ref. \cite{vocadlo} and the $g_{FeFe}$ calculated here. The two are
very similar, and provide evidence that a small percentage of sulfur
impurity does not appreciably affect the properties of the liquid.

The integration of the first peak of the rdf's provides a definition of
the coordination number $N^c_{\alpha \beta}$:
\begin{equation}
N^c_{\alpha \beta} = \rho_\beta \int_0^{r^c_{\alpha \beta}}
4 \pi r^2 g_{\alpha \beta}(r) dr,
\end{equation}
where $r^c_{\alpha \beta}$ is the position of the minimum after the
first peak of $g_{\alpha \beta}$. In pure iron liquid it was found
$N^c_{FeFe}=13.8$ \cite{vocadlo}. In the present case we find
$N^c_{FeFe}=11.2$ (which is essentially $13.8 \times x_{Fe}$, since
the two $g_{FeFe}$'s are practically equal). The integration of
$g_{FeS}$ provides coordination numbers $N^c_{FeS}=2.5$ and
$N^c_{SFe}=10.8$, i.e. each iron atom is surrounded by 2.5 sulfur
atoms, and each sulfur atom by 10.8 iron atoms.  We will comment in
the Discussion section on these numbers.

Two main conclusions can be drawn from these results. The first is
that the system equilibrates quickly: in the RS case there is no
evidence of equilibration time at all, and this means that the
equilibrium configurations of the liquid alloy, when this is built up
distributing impurity atoms in a random way throughout the liquid, are
not much different from those of the pure liquid; the starting
configuration of the CS case has been constructed so that there is an
explicit separation of sulfur from iron, and in this case the system
is not in an equilibrium configuration, but the equilibration time is
very short (of the order of 1 ps), and after that time the random
distribution of the impurities throughout the liquid is restored.  The
second conclusion is more important: there is no evidence of sulfur
clustering or formation of linear chains; rather, a sulfur-sulfur
repulsive tendency is apparent. We will try to explain this effect in the
discussion of the electronic properties of the system in the next
section.

\subsection{Electronic structure}\label{electronic}

The structural behavior of the system can be understood in terms of
the electronic structure. In particular, the interesting quantities
are the relative strengths of the Fe-Fe, Fe-S, and S-S bonds.  In
Fig. \ref{dos} we display the electronic density of states (DOS),
i.e. the total number of electronic states per unit energy, for the
configuration of the RS simulation corresponding at $t=7.9$ ps.
Having in mind a tight binding interpretation of the chemical bonds
among the atoms, it is particularly useful to inspect the local
density of states (LDOS), i.e.  the DOS for each atomic species
decomposed into angular momentum resolved contributions.  The $(l,m)$
angular momentum component of the atom $i$ is the projection onto the
spherical harmonic $(l,m)$ of all the wavefunctions in a sphere of
radius $R$ centered on the atom $i$.  For more details about how the
projections are done see Ref. \cite{kresse3}. The LDOS averaged over
all the atoms of each species in the cell is also displayed in
Fig. \ref{dos}.  The value of the sphere radius $R$ is somewhat
arbitrary. We have used $R=0.8$ \AA ~ for both iron and sulfur, which
is roughly half the minimum distance between the atoms, and thus
should not attribute to one atom possible contributions to the LDOS
deriving from neighboring atoms.

Many features are evident in the DOS; referring all the energies to
the Fermi energy, there is a small peak at $\approx -18$ eV, a
shoulder at $\approx -10$ eV, a main broad peak extending from
$\approx -10$ eV to $\approx 5$ eV and a broad feature well above the
Fermi energy. These features can be easily related to the LDOS shown
in the lower panel of Fig. \ref{dos}. The peak at $\approx -18$ eV is
the S($3s$) level, which is isolated from the rest of the DOS.  The
shoulder at $\approx -10$ eV is mainly due to S($3p$), even if a small
Fe($3d$) contribution is also present. The main peak extending from
$\approx -10$ eV to $\approx 5$ eV is essentially Fe($3d$) and the
feature at $\approx 8$ eV is due to S($3d$) and Fe($3d$). The Fe($4s$)
and Fe($4p$) orbitals are not reported; they are small contributions
to the DOS extending from $\approx -10$ eV to $\approx 15$
eV. Disregarding the S($3s$) level, we can focus our attention on the
S($3p$), S($3d$) and Fe($3d$) bands. The S($3p$) band shows a main
peak at $\approx -10$ eV and a somewhat less intense peak above the
Fermi energy, at $\approx 3$ eV; the Fe($3d$) band has a small
shoulder at the same position as the main S($3p$) peak and extends
well above the Fermi energy.  Since the LDOS depend on the choice of
the sphere radius $R$, we have repeated the same analysis using a
smaller radius, $R=0.6$ \AA. Because of the reduced sphere radius, the
absolute intensity of the peaks is also reduced. This is more true for
the sulfur bands, which are less localized around the nuclei than the
iron $3d$ band. However, the relative intensity of the S($3p$) peaks
at $-10$ and $3$ eV is essentially the same as that for $R=0.8$ \AA.
This fact demonstrates that these two peaks are bonding and
anti-bonding states. We will demonstrate later that they actually
result from a S($3p$)-Fe($3d$) hybridization. Assuming for the moment
that this is the case, we can state that the sulfur-iron bond is
predominantly covalent.  A careful inspection of the LDOS reveals that
part of the sulfur-iron bond is also due to S($3d$)-Fe($3d$)
hybridization.  The bonding between Fe atoms occurs by the well known
mechanism of partial filling of the $3d$-band (this is the 
mechanism emphasized by Friedel's analysis \cite{friedel} of the
cohesive and elastic properties of transition-metal crystals).  The
splitting of the sulfur-iron bonding and anti-bonding levels is a
measure of the strength of the bonds. Since this is larger that the
broadening of the Fe($3d$) band we argue that the Fe-S bond is
stronger than the Fe-Fe one. This is consistent with the different
forms of the rdf's described in the previous section, where it was
evident that the average Fe-S distance is lower than the Fe-Fe one.

In order to disentangle the S-S neighboring effect from the S-Fe one,
we have used the RS simulation to analyze the LDOS of two sulfur atoms
in two different environments. The first one, S$_1$, has been chosen
so as to maximize the number of sulfur atoms in the nearest neighbors
shell, and it has 2 sulfur atoms and 9 iron atoms at a distance less
than 2.5 \AA; while the second atom, S$_2$ has been chosen so that
there are no sulfur atoms within the nearest neighbors shell. In this
way S$_1$ makes bonds with other sulfur atoms, while S$_2$ bonds only
with iron atoms.  In the upper panel of Fig. \ref{ldos1} we display
the LDOS of S$_1$ and S$_2$ for the $3s$ and the $3p$ bands. For the
S$_2$ atom (no sulfur bonds) a sharpening of both the $3s$ and the
$3p$ peaks can be observed, when compared with the averaged LDOS. This
demonstrates that the bonding and the anti-bonding peaks do not result
from a S-S bond, and they are actually due to S-Fe hybridization. The
analysis of the projections onto the S$_1$ atom allows us to infer
about the strength of the S-S bonds. In this case S$_1$ is close to
other two sulfur atoms, and the effect of the S-S orbital overlaps is
evident: there is a nice splitting of the $3s$ level and a
splitting-broadening of the $3p$ level.  Since both the $3s$ and the
$3p$ peaks are far from the Fermi energy, there is no appreciable
energy gain when two sulfur atoms come close. This means that the two
different environments (some sulfur in the first neighbor shell and no
sulfur in the first neighbor shell) are energetically almost
equivalent, so that there is no sulfur-sulfur bond at all.

It is interesting to notice that the $3s$ splitting is larger than the
$3p$ splitting. This is not what one would expect for a couple of
isolated sulfur atoms, since the $3s$ orbitals are more localized than
the $3p$ orbitals in the free atom, and therefore they would overlap
less.  But in the present case this behavior is perfectly consistent
and it is a further demonstration of the sulfur-iron bonding
strength. The $3p$ orbitals are hybridized with the surrounding iron
atoms, while the $3s$ orbitals are well localized on the sulfur atoms,
since they essentially do not interact with iron. Because of this
different spatial distribution, when two sulfur atoms come together
the $3s$ orbitals overlap more effectively than the $3p$ ones, which
are engaged with iron, and this results in the larger splitting
observed in Fig. \ref{ldos1}.

A further evidence of the Fe-Fe and Fe-S bond strength difference can
be inferred by the inspection of the effect of the sulfur neighborhood
on the iron atoms. In the lower panel of Fig. \ref{ldos1} we display
the Fe($3d$) band for two selected iron atoms. The first, Fe$_1$ has 6
Fe and 4 S within a distance of 2.5 \AA, while the second, Fe$_2$, has
11 Fe and 1 S within the same distance.  A comparison of the two bands
clearly shows that Fe$_1$, the iron atom with more sulfur nearest
neighbors, has a broader $3d$ band with respect to Fe$_2$. Since the
band extends across the Fermi level, a broader band results in a
lowering of the energy, and then the Fe-S bond must be stronger than
the Fe-Fe one.  The strength difference of the two Fe-Fe and Fe-S
bonds is expected from the relative extension of the sulfur $3p, 3d$
and the iron $3d$ orbitals: since the sulfur orbitals are less
localized around the nuclei than the iron ones, they overlap with iron
more effectively, leading to a larger broadening of the Fe($3d$) band.

In conclusion, the sulfur-sulfur repulsion evident from the analysis
of the structural properties is not a real repulsion effect, but it is
rather due to the stronger iron-sulfur interaction with respect to the
iron-iron and the sulfur-sulfur ones. The iron atoms want to be as
much coordinated as possible with sulfur atoms, while the
sulfur-sulfur interaction is negligible. The combination of these two
facts produces highly sulfur coordinated iron atoms (compatibly with
the concentration) and isolated sulfur atoms. The consequences for the
transport properties of the liquid will be discussed in the next
section.

\subsection{Dynamics}\label{dynamics}

In the liquid phase the atoms are free to diffuse throughout the whole
volume, and this behavior can be characterized by diffusion
coefficients $D_\alpha$ for the two species of atoms, which are
straightforwardly related to the mean square displacement of the atoms
through the Einstein relation \cite{allen}:
\begin{equation}\label{einstein}
\frac{1}{N_\alpha}\langle \sum_{i=1}^{N_\alpha} |{\bf r}_{\alpha
i}(t_0+t) - {\bf r}_{\alpha i}(t_0)|^2 \rangle \rightarrow 6 D_\alpha
t, ~~ {\rm as} ~~ t \rightarrow \infty,
\end{equation}  
where ${\bf r}_{i \alpha}(t)$ is the vector position at time $t$ of
the $i$-th atom of species $\alpha$, $N_{\alpha}$ is the number of
atoms of species $\alpha$ in the cell, and $\langle \rangle$ means
time average over $t_0$.  In studying the long time behavior of the
mean square displacement, it is convenient to define a time dependent
diffusion coefficient $D_\alpha(t)$:
\begin{equation}\label{tempo}
D_\alpha(t) = \frac{1}{6 t N_\alpha}\langle \sum_{i=1}^{N_\alpha} |{\bf
r}_{\alpha i}(t_0+t) - {\bf r}_{\alpha i}(t_0)|^2 \rangle,
\end{equation}  
which has the property that
\begin{equation}
\lim_{t\rightarrow \infty} D_\alpha(t) = D_\alpha.
\end{equation}  
In Fig. \ref{fig:D} we display the iron and the sulfur diffusion
coefficients calculated using Eq. (\ref{tempo}) for the RS simulation.
The four different panels refer to four different time windows, each
of length 2.0 ps, and starting respectively at $t_{window}$ equal 0,
2.0, 4.0 and 6.0 ps from the beginning of the simulation. That is, for
each window $D_\alpha(t)$ is averaged from $t_0 = t_{window}$ to $t_0
= t_{window} + 2.0$.  We recall again that no equilibration time has
been excluded, so that possible non-equilibrium effects should be
evident from systematic different results in the succession of the
time windows.  The meaningful quantity that has to be extracted from
the pictures is the limit of $D_\alpha(t)$ for large times.  Once
again, there is no evidence of time dependent behavior, and the
diffusivity is approximately the same in all the windows. The
difference that can be appreciated from the different windows is an
estimate of the statistical error on $D_{Fe}$ and $D_{S}$.  From this
data we can estimate $D_{Fe} \approx 0.4-0.6 \times 10^{-4}$ cm$^2$
s$^{-1}$ and $D_{S} \approx 0.4-0.6 \times 10^{-4}$ cm$^2$ s$^{-1}$,
which are very similar. The value of the iron diffusion coefficient is
very close to that found by Vo\v{c}adlo {\it et al.}  \cite{vocadlo} for
pure iron at the same temperature and $\rho = 13.3$ g/cm$^3$, which
was $D_{Fe} \approx 0.4-0.5 \times 10^{-4}$ cm$^2$ s$^{-1}$.

The viscosity of the liquid could in principle also be directly
calculated from the AIMD simulation , {\it via} the autocorrelation
function of the off-diagonal part of the stress tensor \cite{hansen}.
But this would be a major undertaking, and in fact the viscosity has
not yet been calculated for any system by AIMD. The reason is that, by
contrast with the diffusion coefficient, only the average over time
origins can be done in this case, so that the statistics is worse by a
factor $N_{at}$ then that of the diffusion coefficient.  This implies
that for a meaningful measure of the viscosity a much longer run would
be needed.  An alternative way to obtain a rough estimate of the
viscosity is by using its relationship with the diffusion coefficient
stated by the Stokes-Einstein relation:
\begin{equation}\label{stokes}
D\eta = {{k_B T} \over {2\pi a}},
\end{equation}
as was done in our recent calculation on pure liquid iron
\cite{vocadlo,nature}. This relation is exact for the
Brownian motion of a macroscopic particle of diameter $a$ in a
liquid of viscosity $\eta$. The relation is only approximate when
applied to atoms; however, if $a$ is chosen to be the nearest
neighbors distance of the atoms in the solid, Eq. (\ref{stokes}) provide
results which agree within $40 \%$ for a wide range of liquid metals.

In the present case we have two atomic species, each of them with its
own diffusion coefficient. However, iron and sulfur have a similar
atomic radius at this high pressure, and the similar values for the
two diffusion coefficients that we have found are consistent with the
form of the Stokes-Einstein relation, and provide also an indirect
check of its applicability in this particular case.  Since the pure
iron diffusion coefficient \cite{vocadlo,nature} is essentially equal
to that found with the present amount of sulfur impurity, we conclude
that the latter has very little effect on the viscosity of the Earth's
liquid outer core. This means that the value of $\eta \approx
1.3\times 10^{-2}$Pa s obtained in our simulation on pure liquid Fe
\cite{vocadlo,nature} should also be valid for the present Fe-S
mixture.

\section{discussion and conclusions}

We have used first principles calculations based on density functional
theory within GGA for the exchange-correlation energy and ultrasoft
pseudopotentials to simulate a liquid iron-sulfur alloy at Earth's
core conditions ($T=6000$ K, $\rho=12.33$ g/cm$^3$, and a molar
fraction of S = 0.1875). We have found that all atoms are closed
packed, so that the total number of neighbors surrounding each atom is
$\ge 12$. As far as Fe-Fe and Fe-S correlations are concerned, the
distribution of Fe and S atoms is essentially random. But S-S
correlation shows an effective repulsion between S atoms, so that the
probability of finding an S atom in the nearest neighbors shell of a
given S atom is much less than would be obtained with a random
distribution. We have presented strong evidence to show that there is
no tendency whatever for S atoms to form chains.

Our study of the electronic structure shows that the bonding is
predominantly metallic/covalent. Our calculated electronic density of
states demonstrates that S form occupied bonding and unoccupied
anti-bonding states with neighboring Fe atoms. The resulting covalent
S-Fe bond is considerably stronger than the bond between Fe atoms, as
we have seen from the magnitude of the energy splitting between the
bonding and the anti-bonding states. We have argued that the strength
of this bond comes from the large spatial overlap between the S($3p$)
and Fe($3d$) states. By contrast, sulfur atoms do not make bonds
between each other.  The strength of the S-Fe bond compared with the
other two explains the effective repulsion behavior between S atoms:
if two S atoms come together two Fe-S bonds are lost and one Fe-Fe
bond is formed, and the total energy is increased.

Now we come back to the question of the Fe-Fe, Fe-S, and S-S
coordination numbers. If sulfur and iron were equal, and their
distributions random, one would expect $N^c_{FeS}=13.8\times x_{S}
\approx 2.6$ and $N^c_{SFe}=N^c_{FeFe}=11.2$.  Since the Fe-S bond is
stronger than the other two, this should result in higher Fe-S and
S-Fe coordination numbers.  On the contrary, we find two slightly smaller
values, $N^c_{FeS}=2.5$ and $N^c_{SFe}=N^c_{FeFe}=10.8$. However,
since iron and sulfur stay closer than iron and iron (as can be
checked by the inspection of the rdf's), the space left to iron atoms
to surround the sulfur is reduced, and therefore the coordination
number is correspondingly lowered.

Our analysis of the dynamics of the Fe and S atoms shows that the
liquid alloy has essentially the same transport properties as the pure
iron liquid.  We have calculated iron and sulfur diffusion
coefficients which are both of $\approx 0.4-0.6 \times 10^{-4}$ cm$^2$
s$^{-1}$, very similar to that of pure liquid iron, $\approx 0.4-0.5
\times 10^{-4}$ cm$^2$ s$^{-1}$, as calculated in Ref. \cite{vocadlo}.
This also means that the iron-sulfur bonds in the liquid, although
stronger than iron-iron bonds, are not strong enough to form molecules
or polymers, at least at these conditions of pressure and temperature.
Since the diffusion coefficients can be related to the viscosity of
the liquid {\it via} the Stokes-Einstein relation, we conclude that
the sulfur impurity has small effects, if any, on the viscosity of the
Earth's liquid core.

The results discussed in this paper seem at first sight rather
difficult to reconcile with the experimental work of LeBlanc and Secco
\cite{secco}, who found for a Fe-S mixture an anomalous increasing of
the viscosity with pressure.  However, we must point out that the
conditions studied were quite different, so that it is not obvious
that the two works could be compared.  We suggest that a future first
principles investigation for a system with the same conditions of
concentration, pressure and temperature as those of that experimental
work could be interesting.

Finally, we think that a direct first principles calculation of the
viscosity {\it via} the autocorrelation function of the off-diagonal
term of the stress tensor is not completely out of the question,
and we are thinking now to address some effort in this direction.

\section{acknowledgments} 

The work of DA is supported by NERC grant GST/O2/1454. We thank the
High Performance Computing Initiative for allocations of time on the
Cray T3D and T3E at Edinburgh Parallel Computer Centre, these
allocations being provided through the Minerals Consortium and the U.K.
Car-Parrinello Consortium. We thank Dr. G. Kresse and Dr. G. de Wijs
for valuable technical assistance, and Dr. D. Sherman for useful
discussions.

\begin{table}
\caption{Structural parameters for three different FeS crystal
structures calculated using a fit to a Birch-Murnaghan equation of
state.  $\rho_0$ is the equilibrium density, $K$ the bulk modulus and
$K'$ its derivative with respect to pressure.  In the first column we
report FLAPW data \protect \cite{sherman}. In the second and in the
third columns we report our calculations in a spin-restricted and
spin-unrestricted PP formalism.}
\label{tab:solid}
\begin{tabular}{lcccc}
& & FLAPW & PP(spin-unrestricted) & PP(spin-restricted) \\
\tableline
CsCl  & $\rho_0$(g/cm$^3$) & 6.18   &  6.0   &  6.35 \\
         & $K$(GPa)            & 190    &  143   &  191  \\
         & $K'$                & 4.06   &  4.09  &  4.11  \\
& & & & \\
NiAs  & $\rho_0$(g/cm$^3$) & 5.67   &  5.55  &  5.94  \\
         & $K$(GPa)            & 178    &  114   &  176   \\
         & $K'$                & 4.31   &  4.84  &  4.34   \\
& & & & \\
NaCl  & $\rho_0$(g/cm$^3$) & 5.58   &  5.77  &  5.77  \\
         & $K$(GPa)            & 171    &  176   &  176   \\
         & $K'$                & 3.87   &  3.95  &  3.95   \\
\end{tabular}
\end{table}

\begin{figure}
\caption{Comparison between PP spin-restricted and spin-unrestricted
calculations of the energy as a function of the volume for FeS in the
CsCl (B2) and in the NiAs (B8) structures.}\label{fig:spin}
\end{figure}		

\begin{figure}
\caption{Spin-unrestricted calculated energy-volume curves for FeS in
the NiAs(B8) and CsCl(B2) structures. The curves are obtained from a
fit of the data to a Birch-Murnaghan equation of state. FLAPW
calculations \protect \cite{sherman} are reported for
comparison.}\label{fig:solid}
\end{figure}		

\begin{figure}
\caption{Radial distribution functions calculated by averaging over
four successive time windows in the simulation starting with S atoms
in random positions.}\label{fig:grRS}
\end{figure}

\begin{figure}
\caption{Radial distribution functions calculated by averaging over
the first 0.5 ps (left panel), and over 2 ps after 1 ps of
equilibration (right panel) for the simulation started with the sulfur
atoms near each other (CS simulation, see text).}\label{fig:grCS}
\end{figure}		

\begin{figure}
\caption{Iron-iron radial distribution functions calculated in the
pure liquid simulation \protect \cite{vocadlo} and in the present
liquid alloy simulation. }\label{comparison}
\end{figure}		

\begin{figure}
\caption{Total electronic density of states (upper panel) and density
of states for each atomic species decomposed into angular momentum
contributions (lower panel).}\label{dos}
\end{figure}		

\begin{figure}
\caption{LDOS for two selected sulfur atoms (upper panel) and two
selected iron atoms (lower panel). The atoms S$_1$ and S$_2$ have
respectively 2 and 0 sulfur atoms at distance less than 2.5 \AA, and 9
and 11 iron atoms within the same distance.  The atoms Fe$_1$ and
Fe$_2$ have respectively 4 and 1 sulfur atoms at distance less than
2.5 \AA, and 6 and 11 iron atoms within the same distance.}\label{ldos1}
\end{figure}		

\begin{figure}
\caption{Iron and sulfur time dependent diffusion coefficients
calculated using Eq. (\ref{tempo}). The four panels refer to four
different time windows taken to make the time averages (see
text).}\label{fig:D}
\end{figure}

\newpage
\centerline{FIGURE 1}
\bigskip
\centerline{\psfig{figure=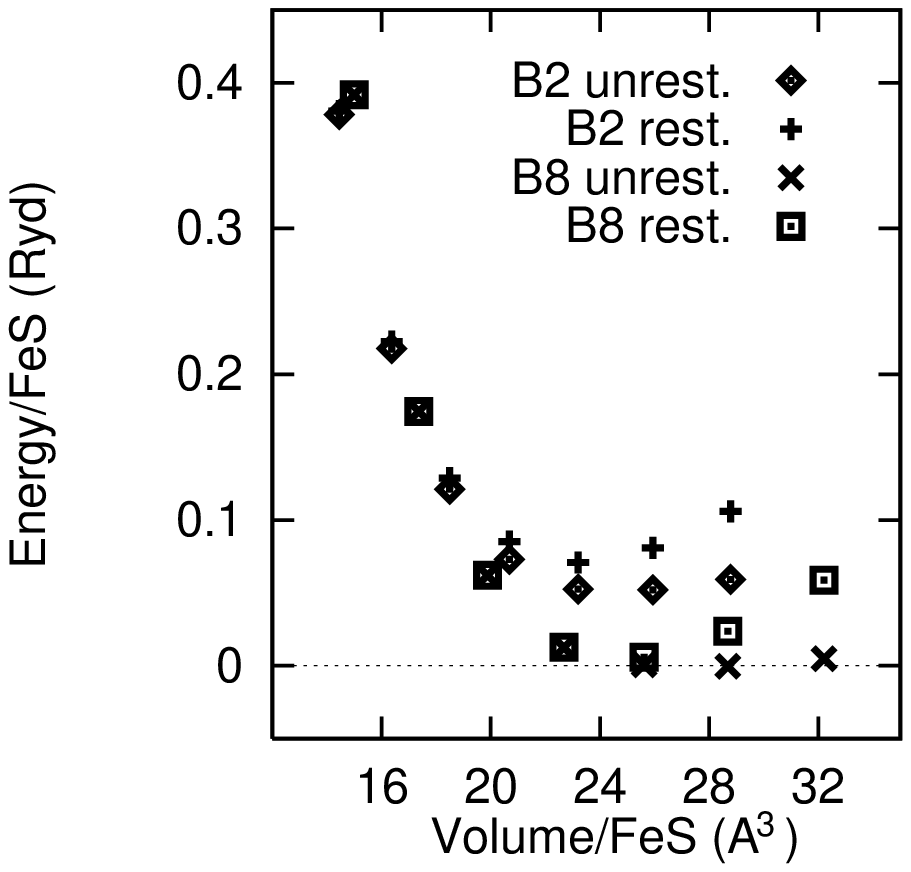,height=3in}} 
\newpage
\centerline{FIGURE 2}
\bigskip
\centerline{\psfig{figure=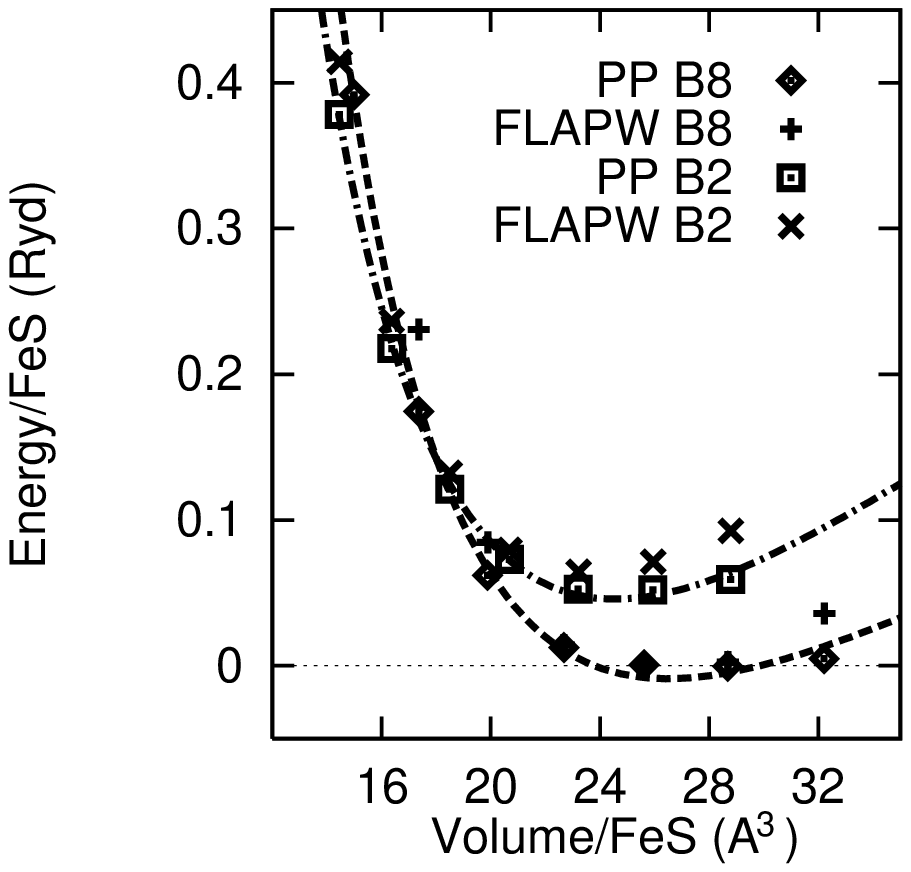,height=3in}} 
\newpage
\centerline{FIGURE 3}
\bigskip
\centerline{\psfig{figure=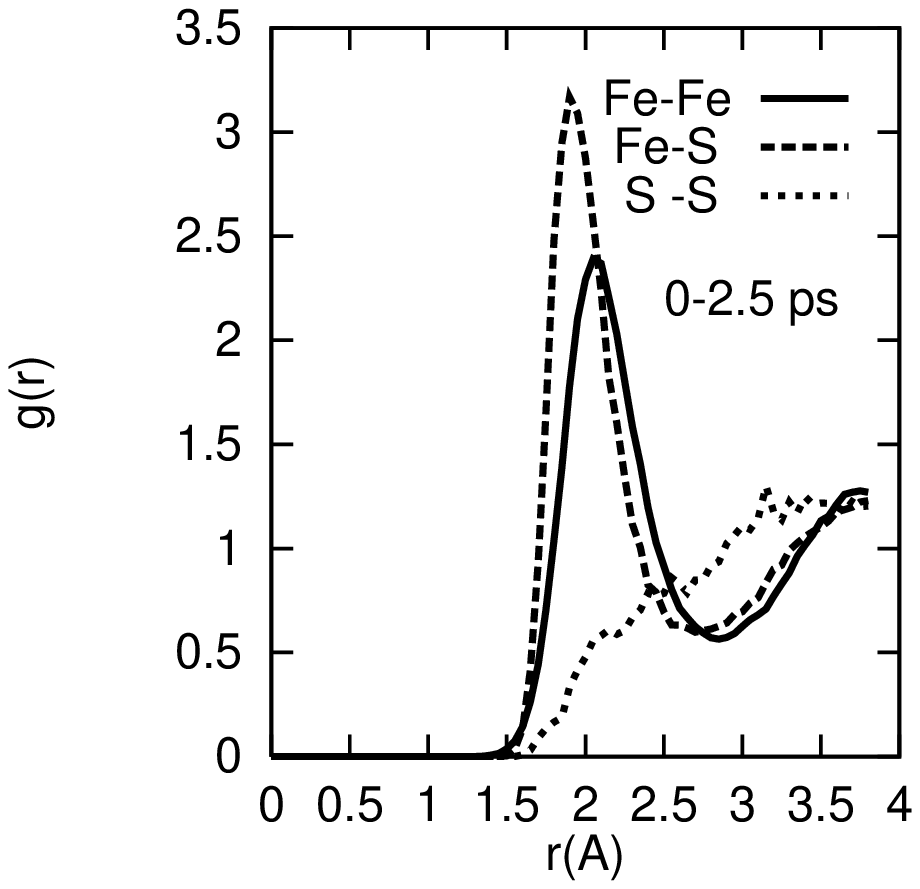,height=3in}
            \psfig{figure=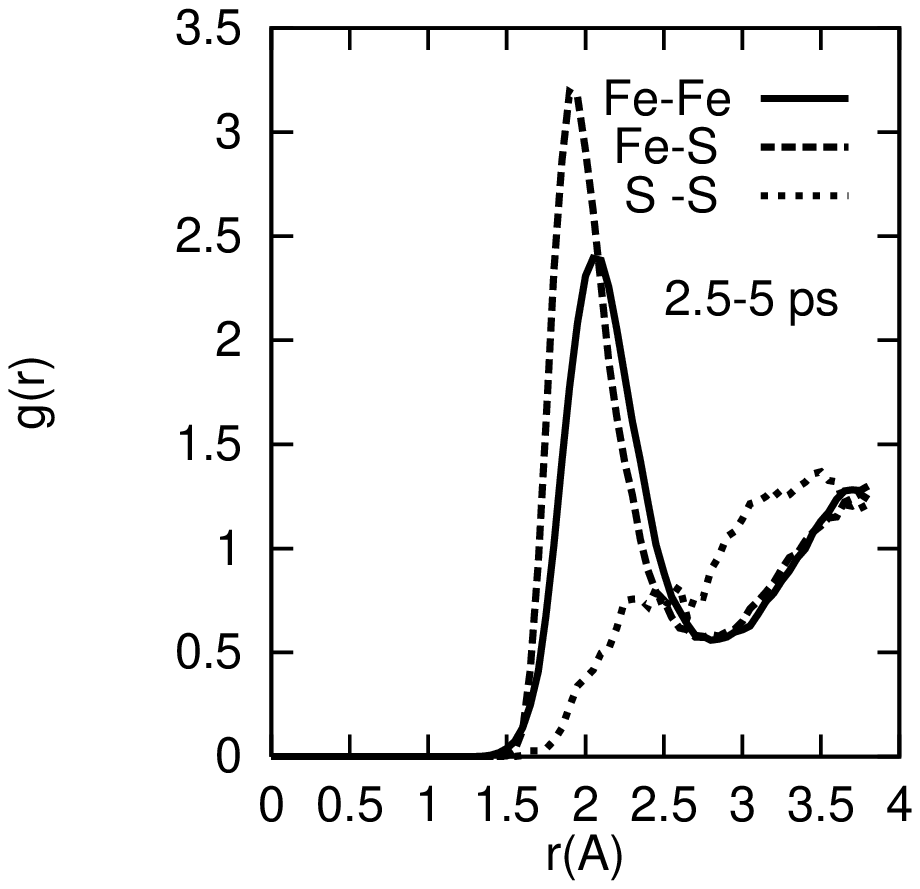,height=3in}} 
\centerline{\psfig{figure=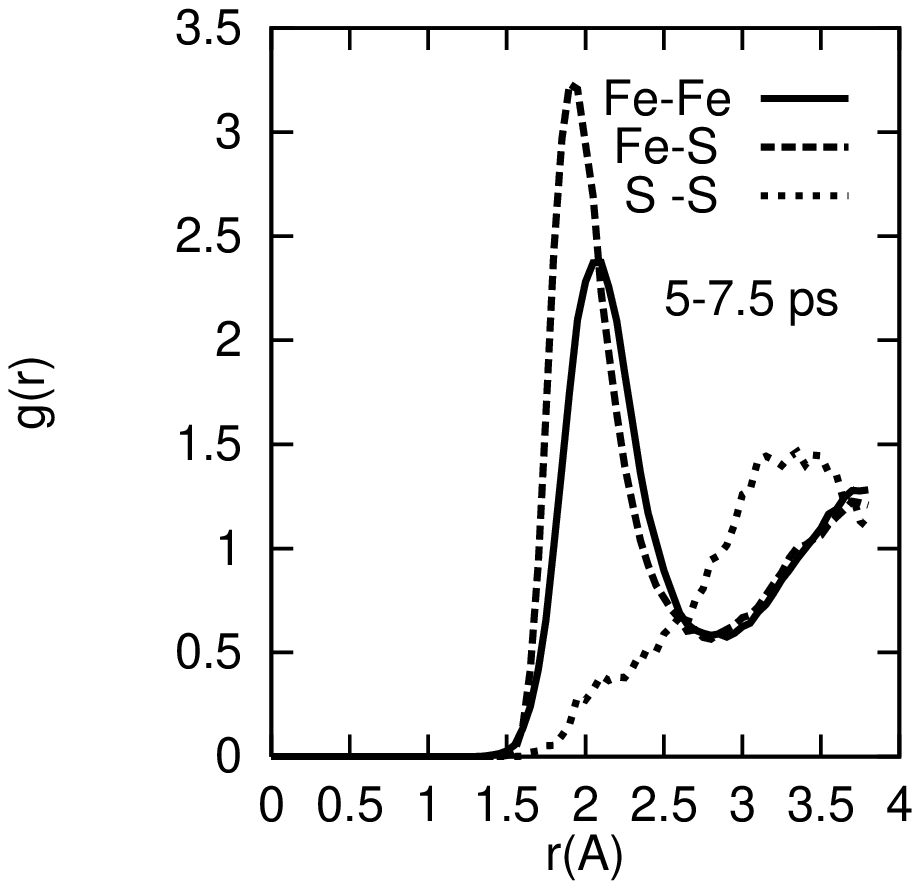,height=3in}
            \psfig{figure=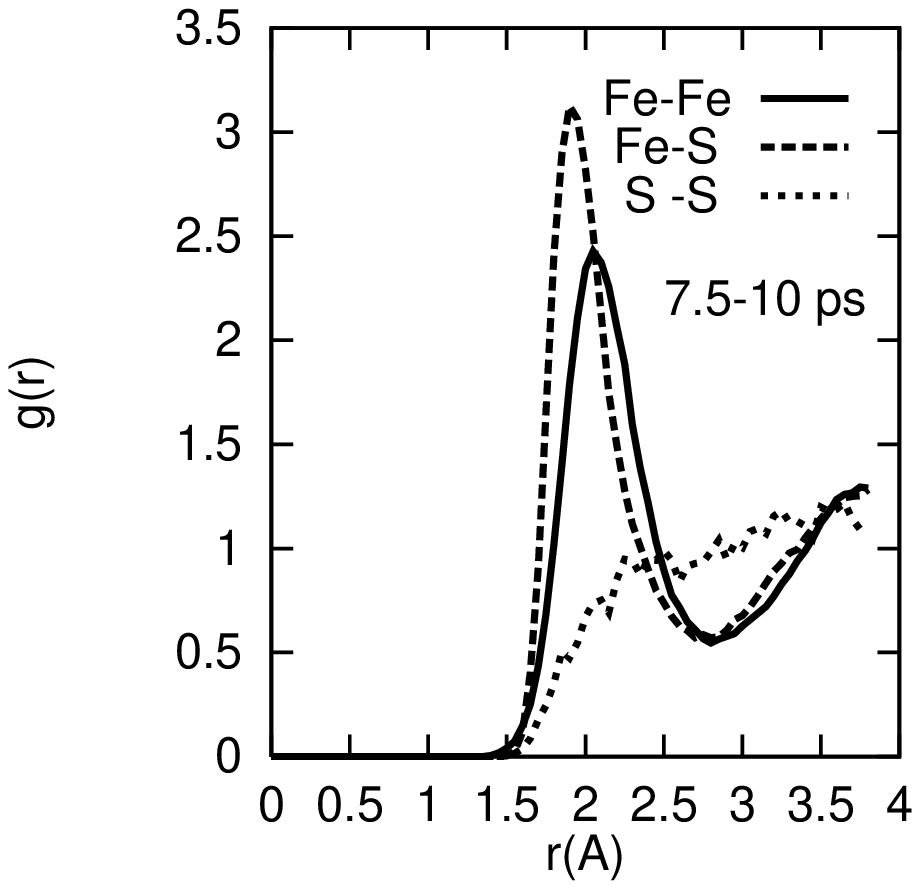,height=3in}} 
\newpage
\centerline{FIGURE 4}
\bigskip
\centerline{\psfig{figure=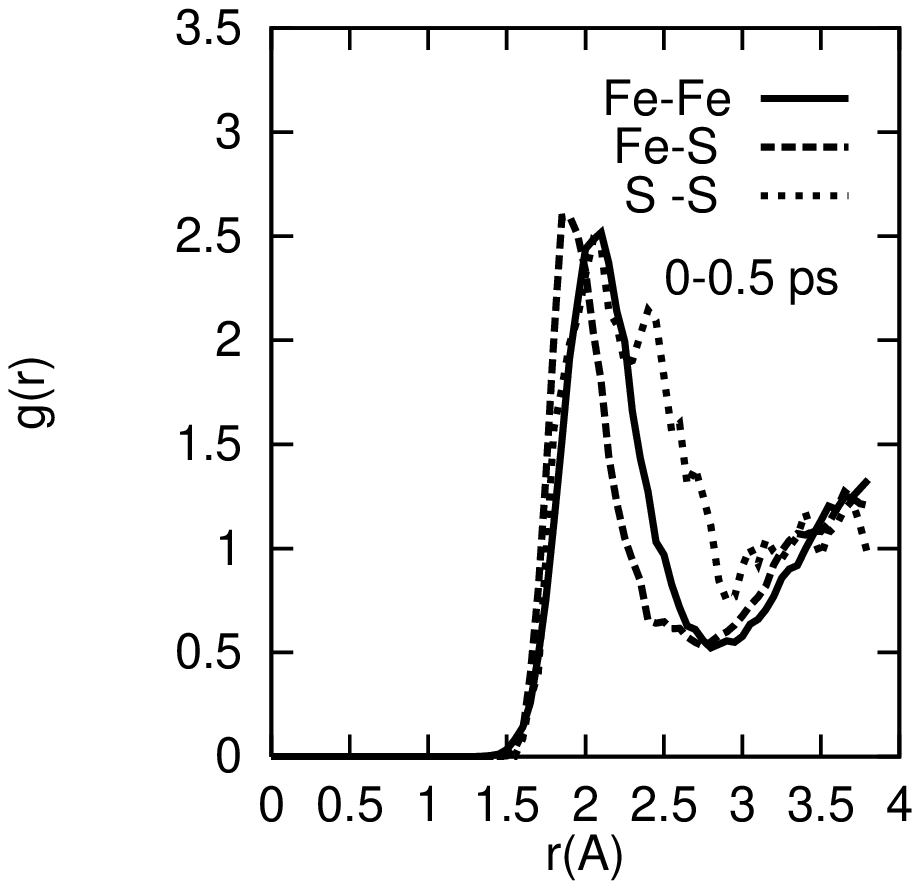,height=3in}
            \psfig{figure=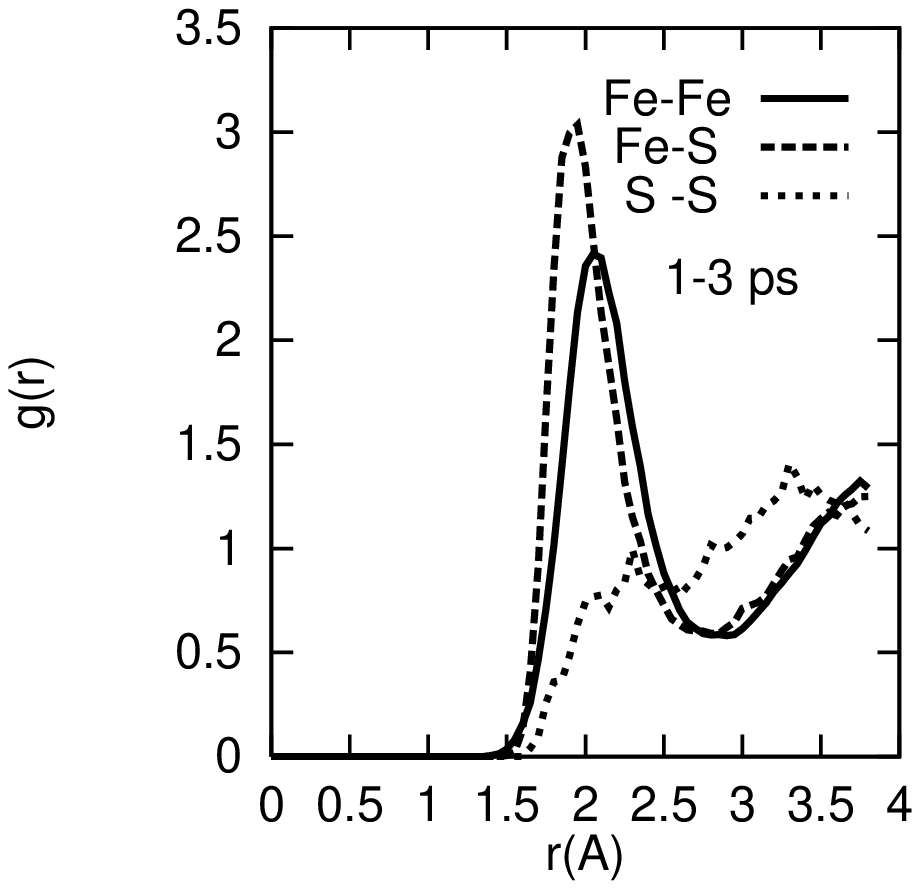,height=3in}} 
\newpage
\centerline{FIGURE 5}
\bigskip
\centerline{\psfig{figure=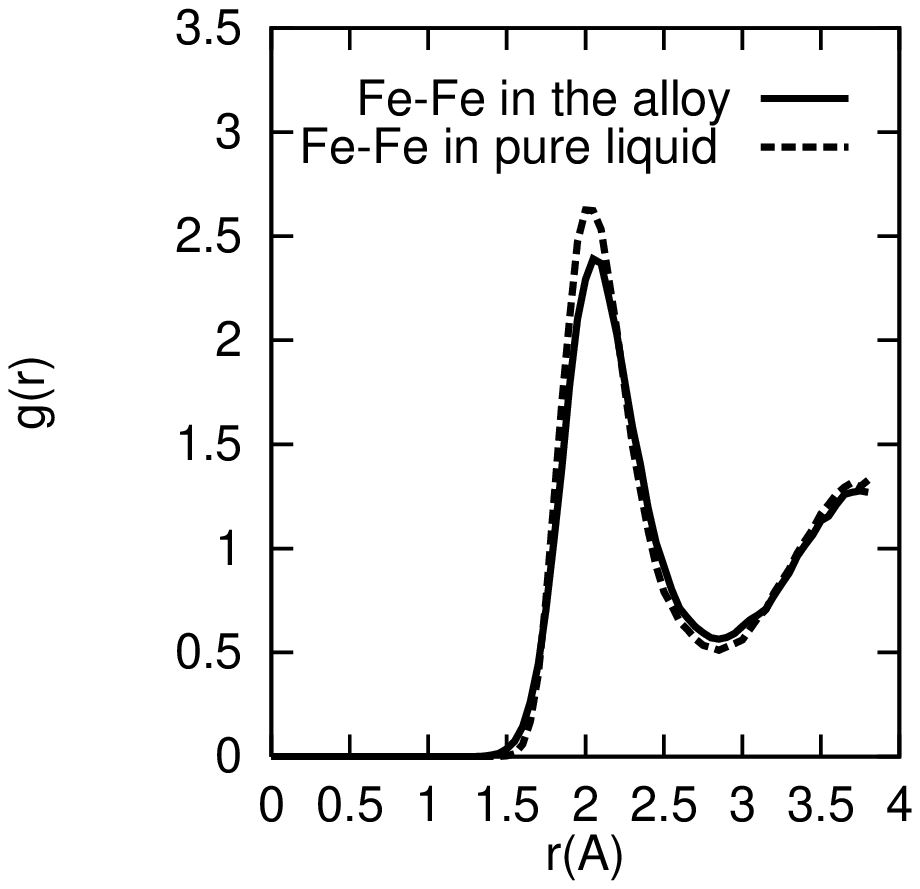,height=3in}}
\newpage
\centerline{FIGURE 6}
\bigskip
\centerline{\psfig{figure=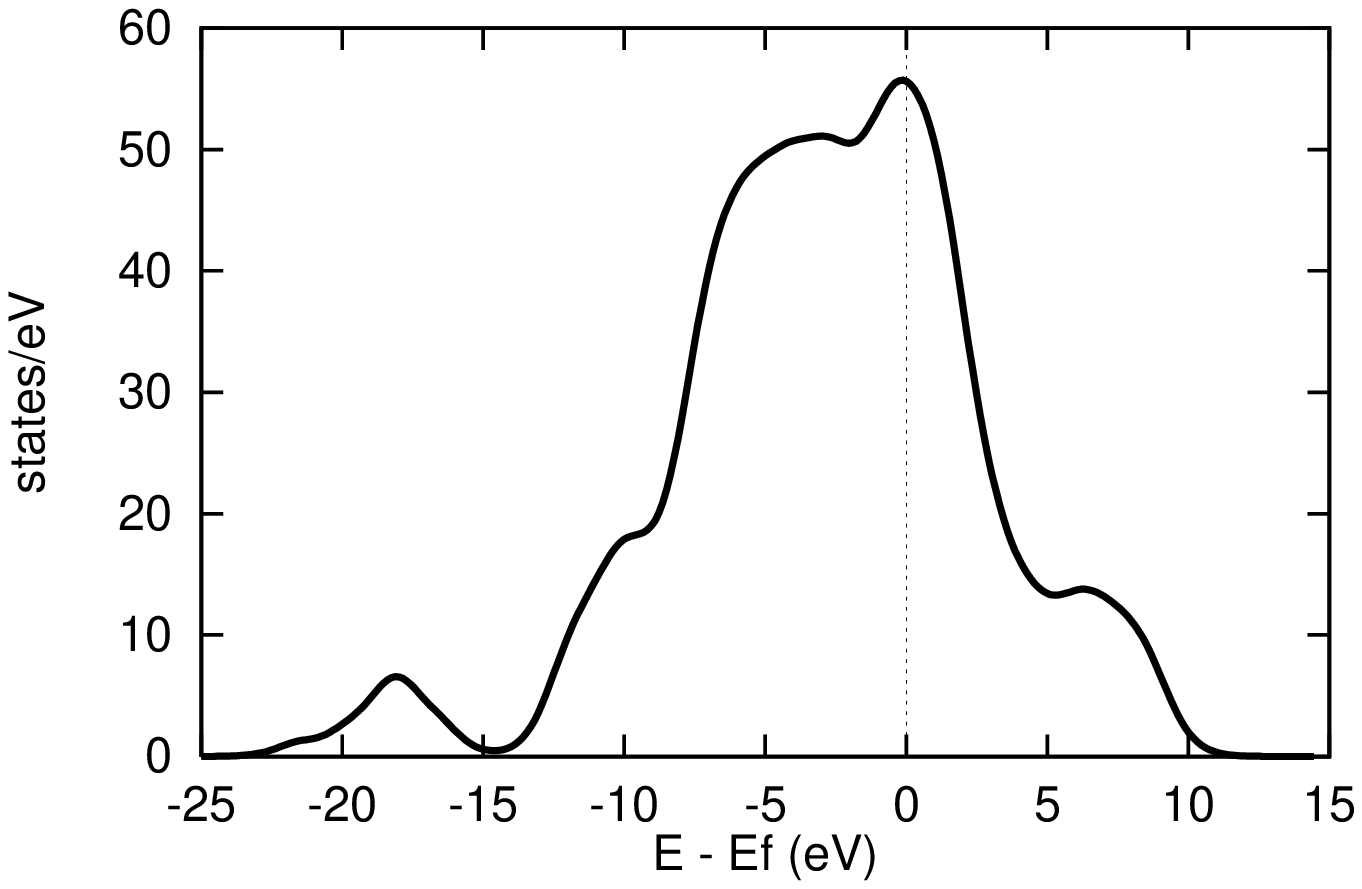,height=3.5in}}
\centerline{\psfig{figure=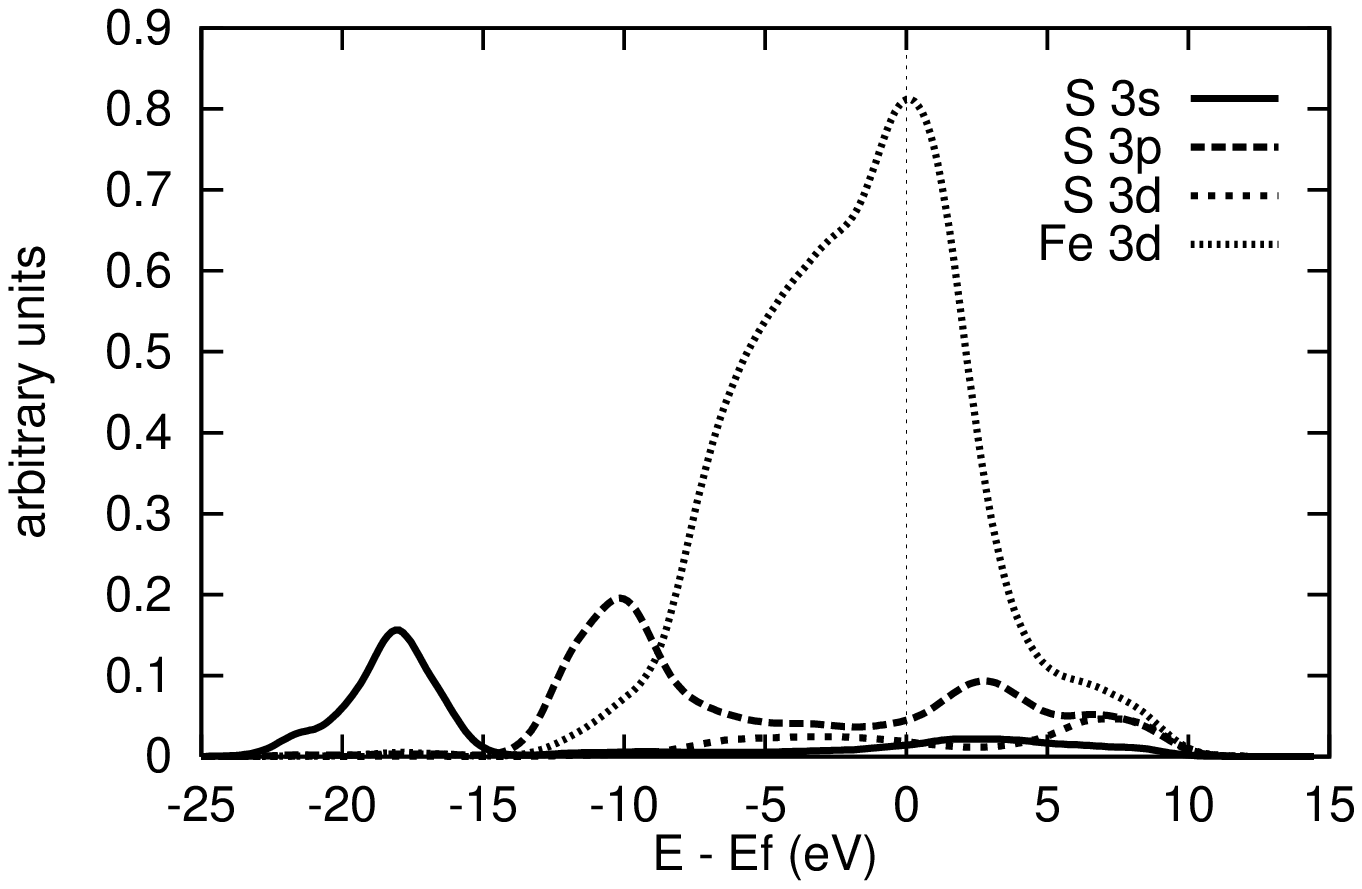,height=3.5in}}
\newpage
\centerline{FIGURE 7}
\bigskip
\centerline{\psfig{figure=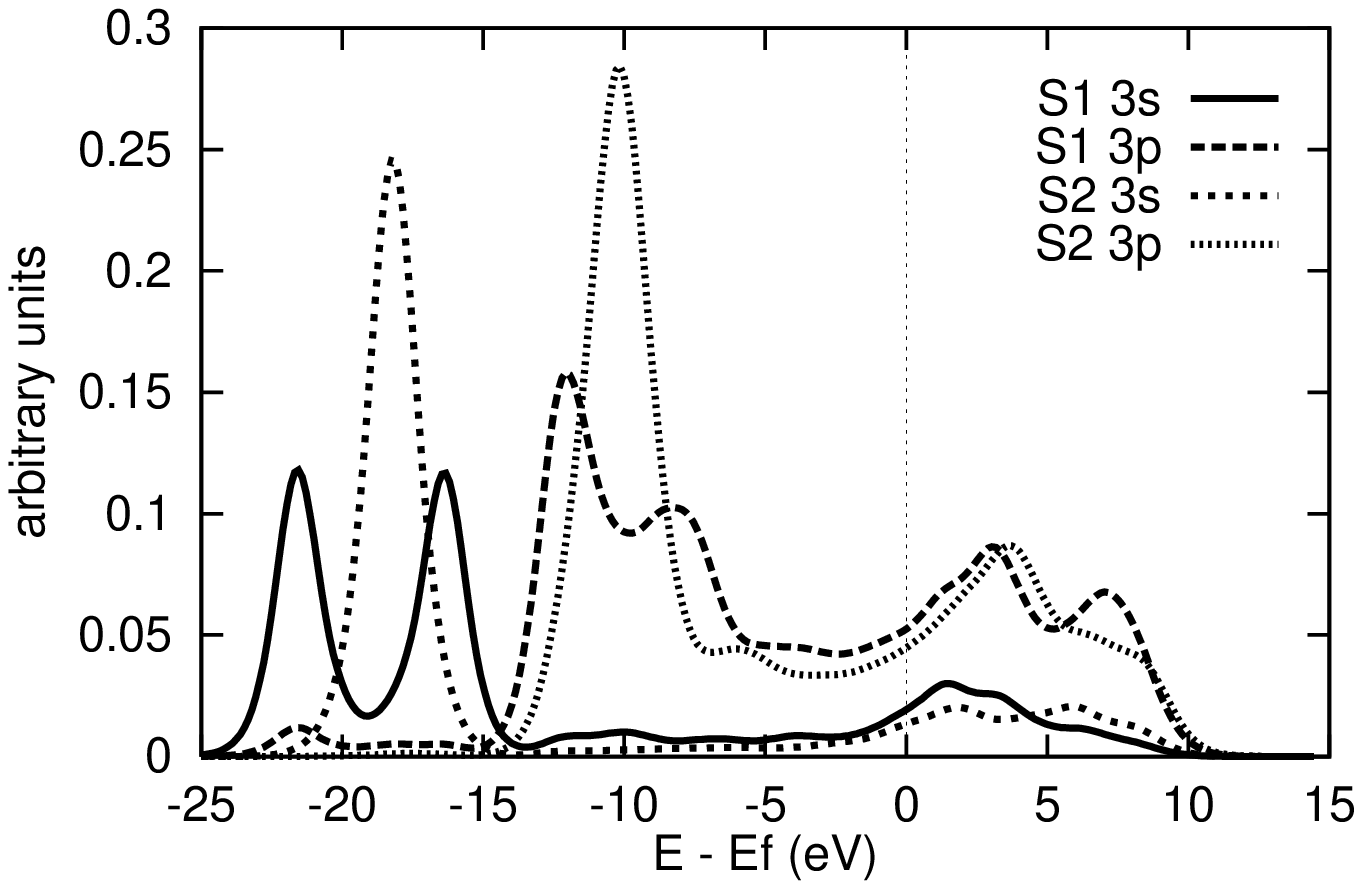,height=3.5in}}
\centerline{\psfig{figure=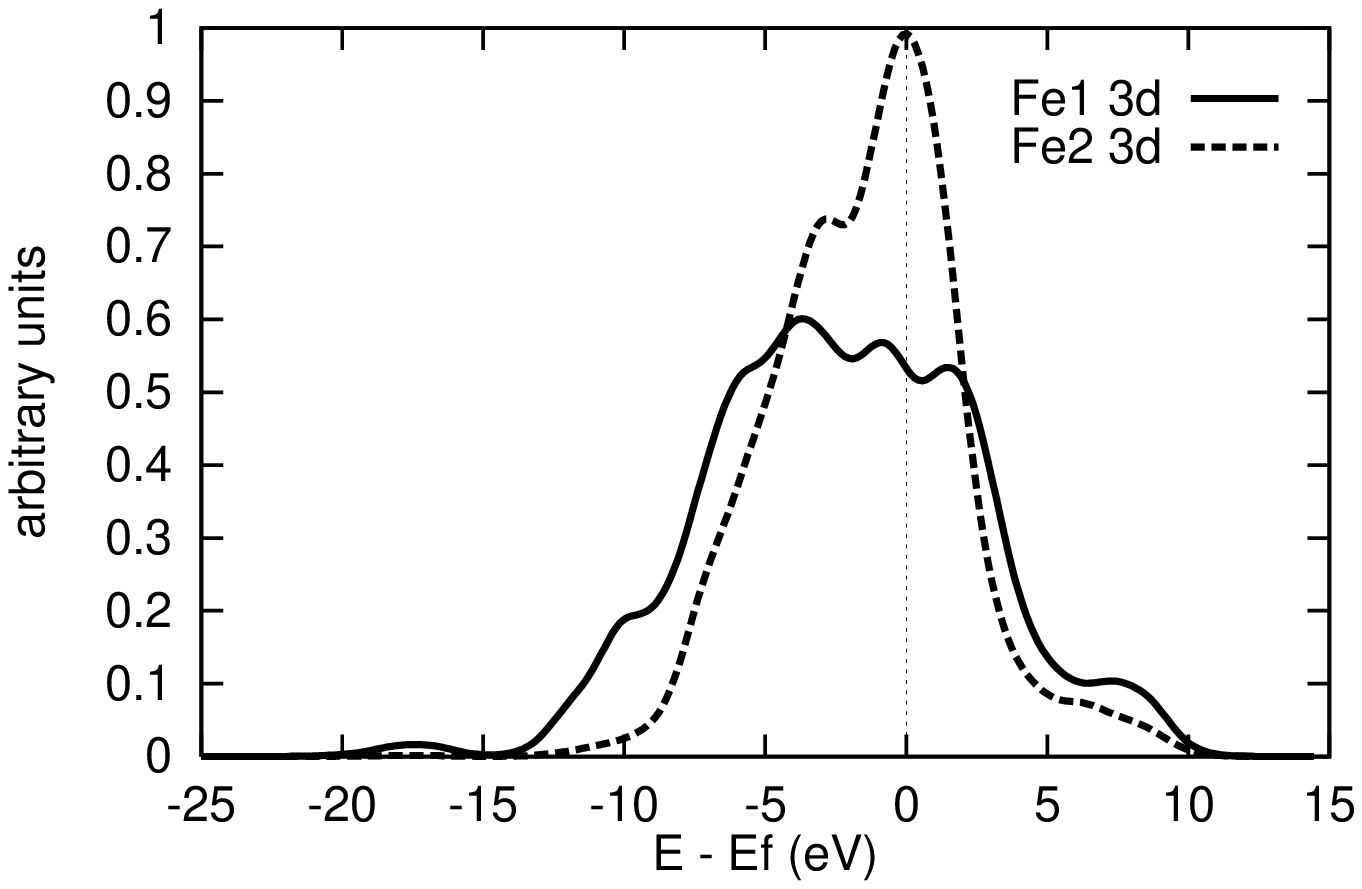,height=3.5in}}
\newpage
\centerline{FIGURE 8}
\bigskip
\centerline{\psfig{figure=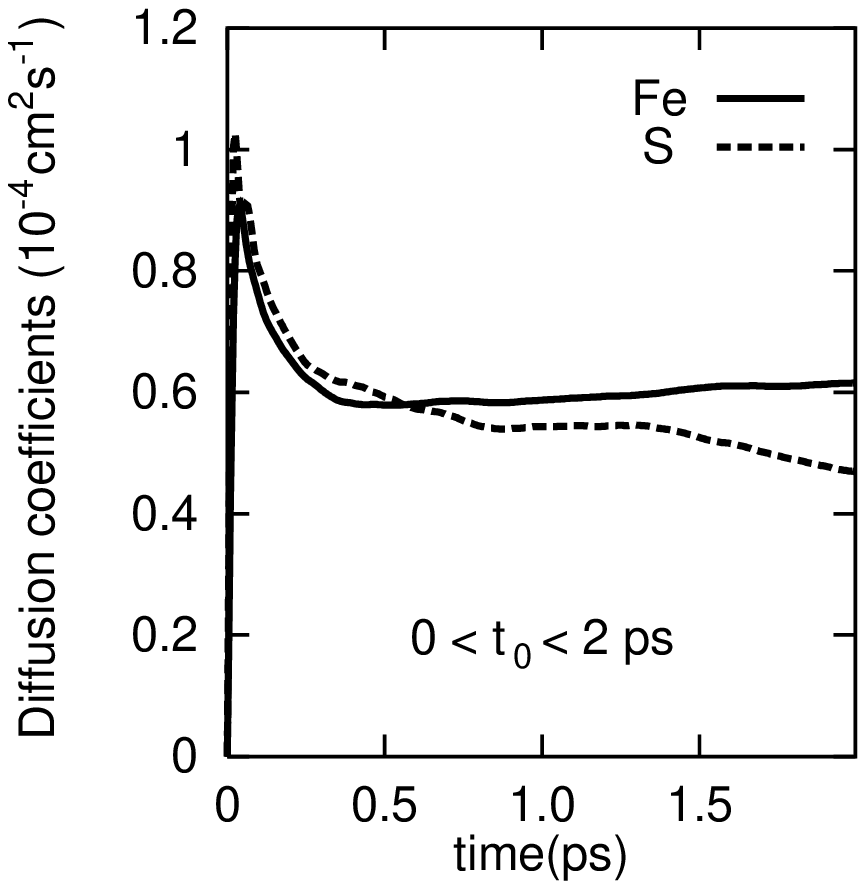,height=3in}
            \psfig{figure=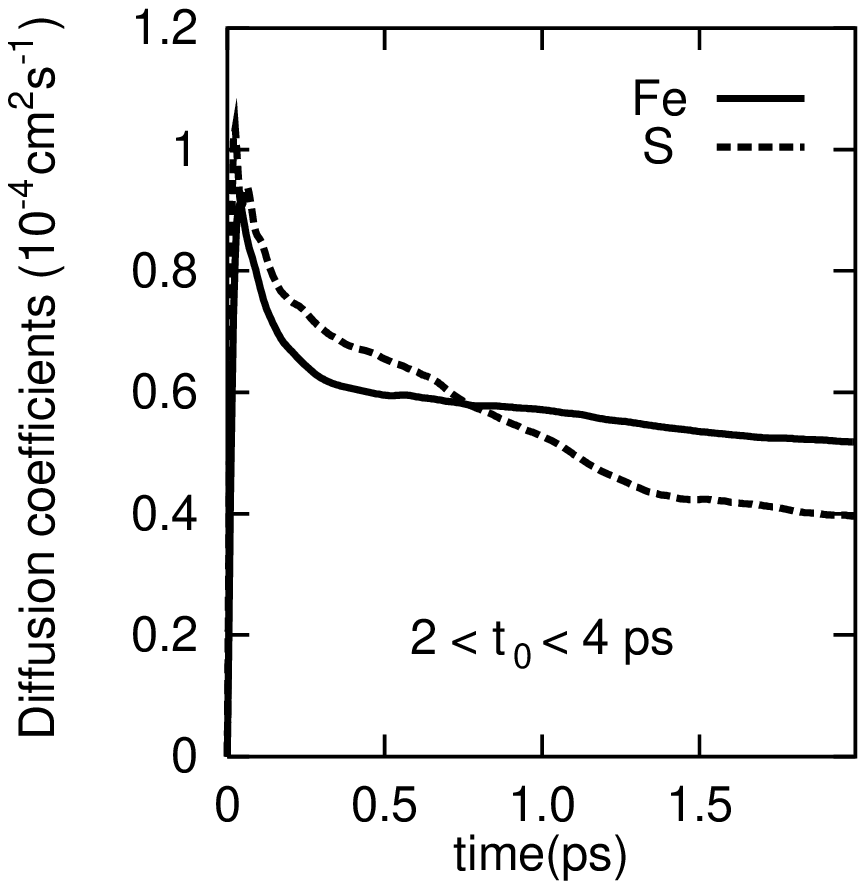,height=3in}} 
\centerline{\psfig{figure=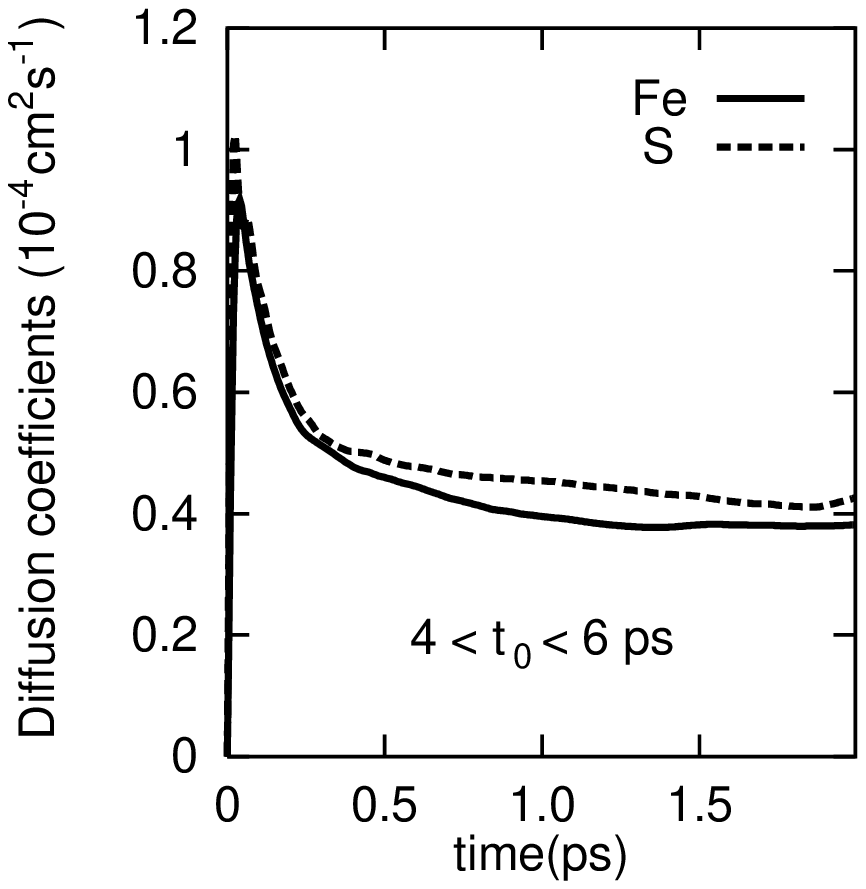,height=3in}
            \psfig{figure=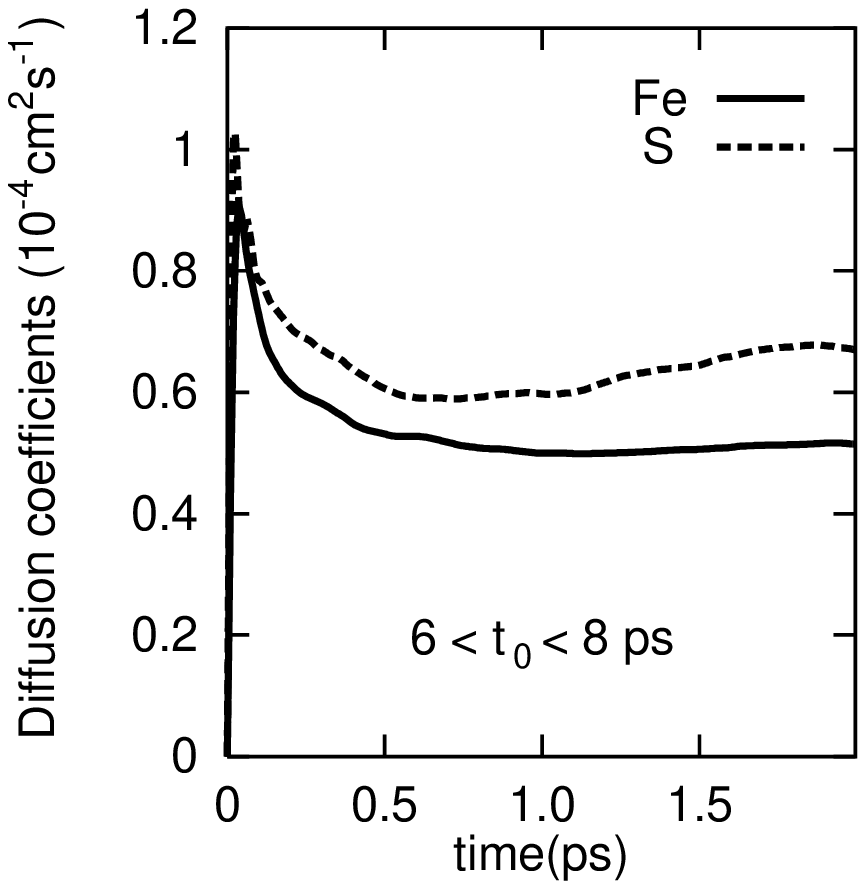,height=3in}}

\end{document}